\documentclass[pdflatex,sn-mathphys-num]{sn-jnl}

\usepackage{graphicx}%
\usepackage{todonotes}
\usepackage{multirow}%
\usepackage{amsmath,amssymb,amsfonts}%
\usepackage{amsthm}%
\usepackage{mathrsfs}%
\usepackage[title]{appendix}%
\usepackage{xcolor}%
\usepackage{textcomp}%
\usepackage{manyfoot}%
\usepackage{booktabs}%
\usepackage{algorithm}%
\usepackage{algorithmicx}%
\usepackage{algpseudocode}%
\usepackage{listings}%
\usepackage{graphicx}
\usepackage{placeins}
\usepackage{doi}

\theoremstyle{thmstyleone}%
\theoremstyle{thmstyletwo}%

\theoremstyle{thmstylethree}%
\newtheorem{definition}{Definition}%

\begin{document}

\title{Per-Shot Evaluation of QAOA on Max-Cut: A Black-Box Implementation Comparison with Goemans–Williamson}


\author*[1]{\fnm{Evgenii} \sur{Dolzhkov}}\email{evgenii.dolzhkov@ut.ee}

\author[2]{\fnm{Franz G.} \sur{Fuchs}}\email{franz.fuchs@sintef.no}

\author[1]{\fnm{Dirk Oliver} \sur{Theis}}\email{dirk.theis@ut.ee}

\affil[1]{\orgdiv{Institute of Computer Science}, \orgname{University of Tartu}, \orgaddress{\street{Narva mnt 18}, \city{Tartu}, \postcode{51009}, \country{Estonia}}}

\affil[2]{\orgdiv{Department of Mathematics and Cybernetics}, \orgname{SINTEF AS}, \orgaddress{\street{Forskningsveien 1}, \city{Oslo}, \postcode{0373}, \country{Norway}}}


\abstract{The Quantum Approximate Optimization Algorithm (QAOA) has emerged as a promising approach for addressing combinatorial optimization problems on near-term quantum hardware. In this work, we conduct an empirical evaluation of QAOA on the Max-Cut problem, using the Goemans–Williamson (GW) algorithm as a classical baseline for comparison. Unlike many prior studies, our methodology treats QAOA implementations as black-box optimizers, relying solely on default parameter settings without manual fine-tuning. We evaluate specific off-the-shelf QAOA implementations under default settings, not the algorithmic potential of QAOA with optimized parameters. This reflects a more realistic use case for end users who may lack the resources or expertise for instance-specific optimization. To facilitate fair and informative evaluation, we construct benchmark instances using well-known graph generation models that emulate practical graph structures, avoiding synthetic constructions tailored to either quantum or classical algorithms. A central component of our analysis is a per-shot statistical framework, which tracks the quality of QAOA outputs as a function of the number of circuit executions. This enables probabilistic comparisons with the GW algorithm by examining when and how frequently QAOA surpasses classical performance baselines such as the GW expectation and lower bound. Our results provide insight into the practical applicability of QAOA for Max-Cut and highlight its current limitations, offering a framework that can guide the assessment and development of future QAOA implementations.}

\keywords{QAOA, Max-Cut, quantum optimization, Goemans–Williamson, benchmarking}



\maketitle

\section{Introduction}\label{sec1}
The \textbf{Max-Cut} problem,
partitioning a graph’s vertices into two sets to maximize the weight of edges crossing between them,
is a canonical NP-hard problem in combinatorial optimization.
It has been widely studied in theoretical computer science and applied optimization due to its relevance in various fields such as statistical physics, circuit design, and network clustering~\cite{goemans1995improved, alon2001constructing, johnson1999theoretician}. Max-Cut has motivated the development of both exact algorithms for small instances and approximation techniques for large-scale graphs~\cite{hsieh2022approximating, dunning2018works}.

Traditional approaches to Max-Cut include semidefinite programming (SDP) relaxations~\cite{goemans1995improved}, heuristic methods such as local search and genetic algorithms~\cite{dunning2018works}, and quantum-inspired algorithms leveraging tensor networks and Ising models~\cite{herrman2021impact}. Recently, the advent of near-term quantum devices has led to significant interest in hybrid quantum-classical optimization methods, particularly the quantum approximate optimization algorithm (QAOA)~\cite{farhi2014quantum, crooks2018performance, farhi2016quantum, hadfield2018quantum}.

QAOA has been studied extensively as a potential candidate for demonstrating quantum advantage in combinatorial optimization~\cite{shaydulin2024evidence, montanezbarrera2024universalqaoaprotocolevidence}. Theoretical analyses indicate that QAOA can achieve performance guarantees for certain classes of graphs~\cite{wurtz2021maxcut, guerreschi2019qaoa}. However, empirical studies have produced mixed results, with some demonstrating competitive or superior performance against classical methods~\cite{harrigan2021quantum, perez2024variational}, while others highlight its limitations, especially at low circuit depths~\cite{marwaha2021local}.

One of the key challenges in applying QAOA effectively is its strong dependence on parameter initialization. Prior research has shown that QAOA's performance, particularly for combinatorial optimization problems such as Max-Cut, is highly sensitive to the choice of variational parameters $(\boldsymbol{\gamma}, \boldsymbol{\beta})$~\cite{farhi2014quantum, boulebnane2021predicting, crooks2018performance}. Poorly chosen parameters can lead to suboptimal solutions, significantly impacting the approximation ratio and overall performance. The optimization of QAOA parameters is a non-trivial task, often requiring extensive classical computational resources to iteratively fine-tune parameters for each problem instance~\cite{wurtz2021maxcut, shaydulin2024evidence}.

To enhance QAOA's effectiveness, recent works have explored various modifications, including hierarchical ansatz designs~\cite{keller2024hierarchical}, improved parameter initialization~\cite{boulebnane2021predicting}, and problem decomposition techniques~\cite{ponce2023graph}. In parallel, studies on benchmarking quantum optimization algorithms have provided insights into their scalability and hardware feasibility~\cite{bucher2024towards, willsch2020benchmarking}. Notably, hybrid quantum-classical approaches that integrate high-performance computing (HPC) with quantum heuristics have emerged as a promising avenue~\cite{patwardhan2024hybrid}.

Despite these advancements, open questions remain regarding the practical advantage of QAOA over classical heuristics. Studies suggest that achieving a quantum speed-up for Max-Cut requires hundreds of qubits~\cite{guerreschi2019qaoa}, which is beyond the current capabilities of Noisy Intermediate-Scale Quantum (NISQ) devices~\cite{preskill2018quantum}. Moreover, recent evaluations indicate that classical approximation techniques may still outperform QAOA under realistic conditions~\cite{marwaha2021local, shaydulin2019evaluating}.

Given the variability in QAOA implementations, e.g. differences in parameter initialization and optimization strategies, it is essential to evaluate not only QAOA as a general framework for Max-Cut but also specific implementations. In this work, we propose a benchmarking scheme designed to systematically assess different QAOA implementations by comparing their performance against state-of-the-art classical alternatives. Whereas most empirical studies report best-achieved or tuned QAOA performance, we focus on user-facing black-box behavior. Furthermore, we introduce a principled approach for selecting and generating problem instances tailored for robust benchmarking.
Our main contributions are as follows.
\begin{itemize}
    \item In Section~\ref{sec:benchmarkingscheme} we formalize a black-box benchmarking framework for comparing methods solving combinatorial optimization problems. The core principle of our approach lies in per-shot percentile tracking.
    \item To systematize a benchmark for Max-Cut that is applicable for small problem sizes, we generate a suite of random graphs for $|V|=29$ and apply the guard described in Section~\ref{sec:datageneration} to ensure that random-sampling produces good cuts only with negligible probability.
    \item Finally, we apply the developed benchmarking framework to compare QAOA with GW. In Section~\ref{sec:results} we show that, under default parameters and realistic shot budgets, fewer than 10\% of QAOA runs achieve the GW expectation, whereas GW itself requires on average fewer than three random hyperplane samples to exceed its own expectation.
\end{itemize}

We start by briefly describing the relevant background.

\section{Background}
This article compares algorithms that approximate solutions to the Max-Cut problem.
Let $G=(V,E)$ be a weighted, undirected graph 
with a positive weights $w_{ij}$ for each edge $(i,j) \in E$.
Consider a set $\mathcal{P}$ of all possible vertex partitions of the kind $P =(S,V\setminus S )$.
We encode a cut through a vector  $\mathbf{x}=(x_1, x_2, \ldots, x_{|V|} ) \in \{-1,1\}^{|V|}$,
where
$x_i = 1$ indicates that the vertex $v_i$ belongs to one side fo the partition and $x_i=-1$ to the other.
The cut value associated with $\mathbf{x}$ is given by the cost function
\begin{equation*}
    C(\mathbf{x}) =\sum_{(i, j) \in E} w_{ij} \frac{1 - x_i x_j}{2}.
\end{equation*}
Thus, the Max-Cut problem can be written as
\begin{equation*}
    \max_{\mathbf{x} \in \{-1,1\}^{|V|}} C(\mathbf{x}).
\end{equation*}
Since the Max-Cut problem is known to be NP-hard, finding the exact solution is infeasible for large graphs.
In practice, it is often enough to have an approximate solution whose quality is measured by the 
approximation ratio defined as
\begin{equation*}
    \alpha_\text{alg} = \frac{C_\text{alg} - C_{\min}}{C_{\max} - C_{\min}} \in [0,1],
\end{equation*}
where $C_{\min}, C_{\max}$ are, respectively, the minimum and maximum possible values of the cost function \(C(x)\).
For Max-Cut with a graph with positive weights, $C_{\min} =0$.
In particular, \(\alpha_{\mathrm{alg}} = 1\) if the algorithm always finds an optimal cut, and \(\alpha_{\mathrm{alg}} = 0\) if it always returns a worst-possible cut.

If the algorithm is probabilistic, $C_\text{alg}$ is defined as the expectation value of the cut.
In order to create a fair comparison, the variance also needs to play a role.
This is in particular true for a quantum algorithm, where it is essential to differentiate between the execution of a single quantum circuit and a complete algorithmic run.
To ensure clarity, the following definitions are introduced.
\begin{definition}[Shot]
A shot refers to a single evaluation of a fixed algorithm applied to an instance. The output of a shot is a bitstring representing a candidate cut, typically sampled from a probability distribution defined by the algorithm, or deterministically computed in the case of non-stochastic methods.
\end{definition}
For QAOA, shots correspond to repeated measurements of a fixed circuit, whereas for GW each shot corresponds to an independent random hyperplane rounding.
\begin{definition}[Run]
A run refers to a complete execution of an algorithm, consisting of one or more shots. Each shot produces a bitstring corresponding to a candidate solution, and the run encompasses all computations required by the algorithm to generate and optionally evaluate these bitstrings.
\end{definition}

\subsection{Classical algorithm for Max-Cut}
An optimal polynomial-time approximation algorithm is the Goemans-Williamson (GW) algorithm, which  
guarantees an approximation ratio of $\alpha_\text{GW} \approx 0.878$ for the Max-Cut problem~\cite{goemans1995improved}.
If the unique games conjecture is true, this is the best possible approximation ratio for Max-Cut~\cite{khot2007optimal}.
The GW algorithm 
consists of the following general three step procedure
\begin{enumerate}
    \item Relax the integer quadratic program into a semi-definite program (SDP).
    \item Solve the SDP to an arbitrarily small error $\epsilon$.
    \item Round the SDP solution to obtain an approximate solution to the original integer quadratic program.
\end{enumerate}
In order to approximate solutions efficiently, a key step in the GW algorithm is to  relax the problem into an SDP form. Let \( \mathbf{X} \) be a matrix such that \( \mathbf{X}_{ij} = x_i x_j \). Since \( x_i^2 = 1 \) for all \( i \), \( \mathbf{X} \) is a positive semidefinite matrix with diagonal entries equal to 1 (\( \mathbf{X}_{ii} = 1 \)). The relaxed SDP formulation is given by:
\begin{equation*}
\max_{\mathbf{X} \succeq 0} \ \frac{1}{2} \sum_{(i, j) \in E} w_{ij} (1 - \mathbf{X}_{ij}) \quad \text{s.t. } \mathbf{X}_{ii} = 1 \quad \forall i \in V.
\end{equation*}
where \( \mathbf{X} \succeq 0 \) indicates that \( \mathbf{X} \) is positive semidefinite.


The rounding procedure for a given optimal solution  \( \mathbf{X}^* \) of the SDP can be done as follows.
Factorize the optimal solution matrix as
$
\mathbf{X}^* = \mathbf{Y}^\top \mathbf{Y},
$
where \( \mathbf{Y} = [\mathbf{y}_1, \mathbf{y}_2, \dots, \mathbf{y}_n] \) is a set of unit vectors in a high-dimensional space.
%
Generate a random vector \( \mathbf{r} \) uniformly distributed on the unit sphere. Each vertex \( v_i \) is assigned to one of the two partitions based on the sign of the inner product:
$
x_i = \text{sgn} \langle\mathbf{y}_i, \mathbf{r} \rangle.
$ 

The expected value of the ratio $\alpha_{\mathrm{alg}}$ produced by the GW algorithm for a weighted graph  $G = (V, E)$ is given by:
\begin{equation*}
\mathbb{E}[\alpha_{\mathrm{alg}}] = \frac{2}{\pi} \sum_{(i, j) \in E} \frac{w_{ij} \cdot \arccos\langle\mathbf{y}_i, \mathbf{y}_j\rangle}{C_{\max} - C_{\min}}.    
\end{equation*}
%

\subsection{Quantum algorithm for Max-Cut}
The quantum approximate optimization algorithm (QAOA) is a hybrid quantum-classical algorithm designed to tackle combinatorial optimization problems on near-term quantum devices. QAOA combines a quantum variational ansatz with classical optimization techniques to approximate solutions to NP-hard problems. QAOA is particularly suitable for problems that can be formulated as finding the ground state of an Ising Hamiltonian. For instance for $3$-regular non-weighted graphs QAOA for Max-Cut at depth $p=1$ guarantees a minimal approximation ratio of at least $0.6924$~\cite{farhi2014quantum}. 

QAOA operates by alternating between two types of unitary operators applied to a quantum state:
\begin{enumerate}
    \item The problem Hamiltonian, \( \hat{H}_C \), encodes the optimization problem. Its ground state corresponds to the optimal solution.
    \item The mixer Hamiltonian, \( \hat{H}_B \), promotes exploration of the solution space.
\end{enumerate}

Given an initial state $|\psi_0\rangle$, the algorithm applies a sequence of alternating unitaries controlled by variational parameters
\begin{equation*}
\psi(\boldsymbol{\gamma}, \boldsymbol{\beta})\rangle = \left(\prod_{k=1}^p e^{-i\beta_k \hat{H}_B} e^{-i\gamma_k \hat{H}_C}\right) |\psi_0\rangle,
\end{equation*}
where  \( \boldsymbol{\gamma} = (\gamma_1, \gamma_2, \dots, \gamma_p) \) and \( \boldsymbol{\beta} = (\beta_1, \beta_2, \dots, \beta_p) \) are variational parameters, and \( p \) is the depth of the circuit.
The expectation value of the problem Hamiltonian is then evaluated via the expectation value
\begin{equation*}
\langle \hat{H}_C \rangle = \langle \psi(\boldsymbol{\gamma}, \boldsymbol{\beta}) | \hat{H}_C | \psi(\boldsymbol{\gamma}, \boldsymbol{\beta}) \rangle.    
\end{equation*}
The goal is to optimize \( \boldsymbol{\gamma} \) and \( \boldsymbol{\beta} \) using a classical optimizer to minimize (or maximize) \( \langle \hat{H}_C \rangle \), thereby finding approximate solutions to the problem.

For the Max-Cut problem, the problem Hamiltonian \( \hat{H}_C \) is given by
\begin{equation*}
\hat{H}_C = \sum_{(i, j) \in E} w_{ij} \frac{1 - \hat{Z}_i \hat{Z}_j}{2},   
\end{equation*}
where \( \hat{Z}_i \) is the Pauli-Z operator acting on qubit \( i \). The term \( 1 - \hat{Z}_i \hat{Z}_j \) applies a different phase depending on the parity of the computational basis state it is applied to.



\section{Methodology}\label{methodology}
This section introduces a generalized framework for benchmarking QAOA in the context of the Max-Cut problem. While the methodology is specifically applied to Max-Cut in this study, the underlying principles are broadly applicable to other combinatorial optimization problems. The proposed scheme provides a structured approach for evaluating QAOA's performance relative to classical algorithms, ensuring reproducibility and comparability across different problem instances, through randomization and taking statistics over enough instances.
A fair comparison of solvers requires three elements: neutrality to specific configurations, unbiased dataset selection, and benchmarking grounded in robust statistical analysis. These criteria are detailed below.

\subsection{Algorithm implementation settings}
The goal of this benchmarking framework is to evaluate the performance of numerical solvers for combinatorial problems under default settings, i.e. without starting parameter tuning.
Most end users do not have deep expertise in parameter tuning; they simply want a workable solution without manually searching for the ``best'' parameters such as error settings for SDP solvers or optimal circuit depth for QAOA.
Therefore, we run each implementation using its default parameters, unless the developers explicitly recommend alternative defaults for specific problem classes.

Using defaults aligns with real‐world usage: if finding good starting angles requires a time‐consuming search, any quantum advantage may vanish once the classical overhead is added. Alternatively, this needs to be part of the run time analysis.
In other words, the promise of QAOA as an efficient optimizer might break down if extensive parameter sweeps are needed to beat a classical solver. If extensive manual parameter search is required to outperform GW, that search cost must be included; we deliberately exclude it here.
By evaluating each codebase in its default configuration, we measure performance in the way most users actually experience it.
Under this setup:
\begin{itemize}
  \item Starting variational parameters $(\gamma,\beta)$ remain at the implementation's provided defaults.
  \item Starting circuit depth $p$ is set to the default (or to the value recommended by developers for the problem type).
  \item If an implementation offers multiple initialization strategies or adaptive depth‐selection heuristics, we adopt the developer's own recommendations for example instances (otherwise, we do not tweak anything).
  \item All quantum simulations are noiseless (i.e., an ideal simulator) since the scope of this scheme is to explore an algorithm implementation's potential without relying on hardware specifics.
  \item Also classical solvers remain with their default parameter settings.
\end{itemize}
This allows us to systematically evaluate the performance of specific implementations of QAOA in solving combinatorial optimization problems with respect to classical alternatives.

\subsection{Fair dataset selection}
\label{sec:datageneration}
Benchmark instances should, where possible, be drawn from established combinatorial‐optimization repositories, such as the Gset library\footnote{\url{https://web.stanford.edu/~yyye/yyye/Gset/}}, which contain Max‐Cut problems that are widely used as benchmarks in the literature. These standardized datasets ensure that results remain directly comparable to prior studies. However, two practical issues frequently arise: (1) many Gset graphs have very large vertex counts, rendering QAOA simulations infeasible, and (2) they often include both positive and negative edge weights, complicating quantum‐algorithm implementations.

To mitigate these challenges, one can generate random graph instances that are (a) sufficiently difficult for the GW relaxation and (b) small enough to allow tractable QAOA simulation. In practice, one might choose a well‐studied random‐graph model (e.g., Erdős–Rényi \(G(n,p)\) or random \(d\)‐regular graphs\footnote{Although, one has to be extra careful when selecting graphs with certain mathematical properties, as they can skew performances of either classical or quantum algorithms. By defaults, it's recommended to select instances that are ``as generic as possible'' unless the scope of the benchmark is deliberately narrowed down to a specific class of graphs with pre-determined properties.}) and then impose three constraints:

\begin{enumerate}
    \item \textbf{GW Expectation Threshold.} Ensure that the expected cut value from the GW relaxation remains below a fixed fraction of the true Max‐Cut value. This prevents the creation of trivial instances, for which GW already achieves near‐optimal performance due to small size or low density, and guarantees that any algorithm must meaningfully improve on the GW bound.
    \item \textbf{Random‐Sampling Hardness.} Require that a uniformly random cut exceed the GW expectation only with negligible probability. Concretely, compute the \((100-\epsilon)\)\nobreakdash-th percentile of the distribution of all possible cut values, for a suitably small \(\epsilon\) (e.g., \(\epsilon = 10^{-8}\) or \(\epsilon = e^{-|V|}\), where $|V|$ is the number of vertices). If even this high percentile remains well below the desired threshold for a ``good'' cut, then any random sampling strategy has at most \(\epsilon\) chance of yielding a near‐optimal solution. Demonstrating that the \((100-\epsilon)\)\nobreakdash-th percentile is low thus rigorously shows that random cuts exceed the GW expectation with probability at most $\epsilon$, which rules out random sampling as a competitive baseline.
    \item \textbf{Sufficient Number of Cuts Exceeding the GW Expectation.}
    Ensuring that the $(100-\epsilon)$-th percentile of the cut value distribution is sufficiently low provides only an upper bound on the number of cuts exceeding the GW expectation. For meaningful benchmarking, however, it is necessary to ensure that each instance contains a sufficiently large set of ``good'' solutions. For example, an instance in which only a single cut --- namely, the Max-Cut --- exceeds the GW expectation is unsuitable for this benchmarking scheme, as it does not allow for a meaningful assessment of how frequently a quantum algorithm identifies such solutions or whether its performance converges to a stable approximation ratio that exceeds the GW expectation while remaining below the optimal value. For graphs with a small number of vertices (i.e., $|V| \leq 30$), the number of cuts exceeding the GW expectation can be computed exactly. For such instances, it is therefore recommended to impose a constraint requiring at least $|V|$ distinct cuts whose values exceed the GW expectation.

\end{enumerate}

Finally, to calibrate solver performance correctly, it is essential to know, or at least closely approximate, the exact Max‐Cut value for each instance. For small graphs, one can use brute-force enumeration to compute the optimal cut. Alternatively, one can generate structured instances with a known Max-Cut by construction. If neither approach is practical, a high-quality heuristic or approximation algorithm (e.g., branch-and-bound, semidefinite-programming relaxations, or specialized Max-Cut solvers) should be used to estimate the optimum. This ensures that all reported performance metrics (approximation ratios, absolute gaps, etc.) are normalized against a reliable ground-truth value.

\subsection{Robust statistical analysis}
\label{sec:benchmarkingscheme}
To analyze probabilistic solvers properly, the output distribution should be tracked on a per-shot basis. Specifically, for a given graph instance with \( n \) vertices, the first \( N \leq 2^{n-1} \) outputs generated should be recorded. The total number of shots per run can be adjusted based on computational resources and experimental preferences and constraints. For instance QAOA execution should be repeated across \( R \) independent runs for each graph instance where within each run, the first \( N \) shots should be recorded. To aggregate shot-by-shot statistics, the best-performing solutions observed at each shot across all independent runs should be analyzed. For a given shot number \( s \), the maximal cut value \(C_{\max}^{(s)}\) encountered among the current shot and all the previous ones is computed as
\begin{equation*}
C_{\max}^{(s)} = \max_{r \in \{1, 2, \dots, s\}} C^{(r)},   
\end{equation*}
where \(C^{(r)}\) denotes the cut value obtained at shot $r$.

After collecting the per-shot maxima, the empirical  \((100-\epsilon)\)\nobreakdash-th percentile for all shot indices \( s \in \{1, 2, \dots, N\} \) should be computed to assess the distribution of high-quality solutions as a function of shot index
\begin{equation*}
P_{100-\epsilon}(s) =   \text{\((100-\epsilon)\)\nobreakdash-th percentile of } \{C_{\max}^{(s)} \text{ across all $R$ runs}\}.    
\end{equation*}

Once the empirical percentiles have been computed, they can be compared against the GW lower bound and GW expectation to determine the probability of QAOA surpassing classical performance thresholds.

This best-so-far statistical framework is consistent with the operational structure of QAOA, wherein the classical post-processing stage selects the highest-quality sample obtained among all shots within a given run. Tracking the cumulative best value as a function of the shot index therefore provides a faithful representation of the algorithm’s practical behavior and enables performance assessment up to any prescribed shot budget. Moreover, this evaluation methodology corresponds to an optimistic yet operationally justified interpretation of shot-based QAOA performance, as it reflects the best solution that would realistically be retained during execution.

\section{Test results} \label{sec:results}
This section reports the empirical evaluation of the black‑box QAOA implementation described in
Section~\ref{methodology} against the GW classical baseline.  
Results are organized around three guiding questions:
\begin{enumerate}
  \item How quickly does QAOA produce a cut that beats the GW lower bound?
  \item With given shot budgets, what fraction of QAOA runs reach or exceed the GW expectation?
  \item How many random‑hyperplane samples does GW need to reach the same quality levels?
\end{enumerate}

All QAOA runs use fixed parameters throughout each run. Unless otherwise stated, all numerical settings --- graph size $\lvert V\rvert = 29$, run count $R = 1,000$, shot budget $N = \lfloor 2^{\lvert V\rvert / 2} \rfloor$, GW sample budget $K = 10^{8}$ (see Appendix~\ref{appendix:gwstats}), and instance-generation restrictions --- follow Appendix~\ref{appendix:generatingrestrictions}.

For each instance, we plot QAOA performance relative to both the GW lower bound and the GW expected value. Plots also include the percentage of QAOA runs surpassing each of these thresholds, all shown as functions of the shot count.

The specific QAOA implementation~\cite{fuchs2024qaoa} used in this study, along with the benchmarking code\footnote{\url{https://github.com/OpenQuantumComputing/maxcut-classical-quantum-benchmark}} and the instance library, are publicly available. In this paper we present a full list of plots for a single instance ba\_n-29\_m-9\_132 (see Appendix~\ref{appendix:experiments}). The rest of the instances follow a similar pattern of behavior and their respective plots can be found in the benchmarking repository. Instance generation and brute-force Max-Cut computations were performed on the University of Tartu HPC cluster; QAOA execution and data processing were run on an Nvidia H100 GPU.

\subsection{Shot‑resolved convergence of QAOA}

Figures~\ref{fig:90maxcut} and~\ref{fig:99maxcut} contain the cumulative percentage of runs in which \emph{at least one} QAOA shot has already produced a cut surpassing (i)~the GW lower bound and $0.9\cdot C_{\max}$ and (ii)~the GW expectation and $0.99\cdot C_{\max}$.  Across the entire dataset three qualitative patterns emerge:

\begin{enumerate}
\item \textbf{Rapid pass of the lower bound.}  For every instance, most runs exceed the GW lower bound within the first $2,000$ shots (with the exceptions cws\_n-29\_k-6\_p-0.412\_148 and er\_n-29\_p-0.227\_123 that require around $6\,000$ shots), indicating that default‑parameter QAOA relatively quickly leaves the regime of trivially bad cuts.
\item \textbf{Stagnation between the lower bound and the GW expectation.}  Even after the full $N\approx 23\,000$ shots, fewer than $1\%$ of runs exceed the GW expectation, except for ba\_n-29\_m-6\_239, where this number reaches $1.2\%$; in the worst case (instance er\_n-29\_p-0.493\_195) the fraction never rises above $0.2\%$. No instance shows convergence toward the GW expectation as shot count increases.
\item \textbf{Extremely low probability of reaching near-optimal cuts.}  Only a single QAOA run among all runs over the entire dataset (i.e. $7\,000$ runs total) produced a cut greater than $99\%$ of the Max-Cut. 
\end{enumerate}

These observations are quantified in Figure~\ref{fig:ba_n-29_m-9_132_gw}, which tracks the empirical $90$‑th and $99$‑th percentiles $P_{90}(s)$ and $P_{99}(s)$ of the best‑so‑far cut value over all runs (confidence intervals for $90$-th and $99$-th percentiles are obtained via non-parametric bootstrapping). Overall, for no instance does $P_{90}(N)$ reach the GW expectation; in three instances even $P_{99}(N)$ remains below it, confirming that exceptional ``lucky'' shots do not translate into consistent progress across runs.

\subsection{Efficiency of GW hyperplane sampling}
To contextualize QAOA’s shot cost, Figure~\ref{fig:ba_n-29_m-9_132_gw_classical} reports the \emph{expected number of random hyperplanes} required by the GW rounding procedure to achieve specified approximation ratios. Across all graphs:
\begin{itemize}
  \item \mbox{$\le 3$} samples suffice on average to beat the GW expectation;
  \item \mbox{$\le 15$} samples suffice on average to find an optimal cut ($C_{\max}$);
  \item only two instances (cws\_n-29\_k-6\_p-0.229\_199 and ba\_n-29\_m-6\_239) fail to hit $C_{\max}$ within $10^{8}$ samples, yet still reach $\alpha\ge 0.975$.
\end{itemize}
Comparing costs therefore yields a striking asymmetry: GW reaches its own expectation roughly after 3 samplings, whereas for QAOA, convergence to the GW expectation or above is not even observed within a given shot budget.

\subsection{Aggregate comparison and variability}

While individual instances exhibit some variability in convergence speed and success rates, none of these fluctuations alter the overarching conclusion: \emph{with default parameters and without problem-specific optimization}, the evaluated black-box QAOA implementation remains statistically unlikely to surpass even the \emph{expected} performance of the classical GW heuristic within practical shot budgets.

This conclusion is robust across all graph families and parameter settings examined. In particular, the empirical percentiles and instance-level trajectories show no systematic trend toward closing the gap to the GW expectation as shot counts increase. The rare “lucky” high-value cuts that do appear do not accumulate sufficiently across runs to meaningfully raise the typical QAOA performance.

\begin{figure}[h]
    \includegraphics[width=\textwidth]{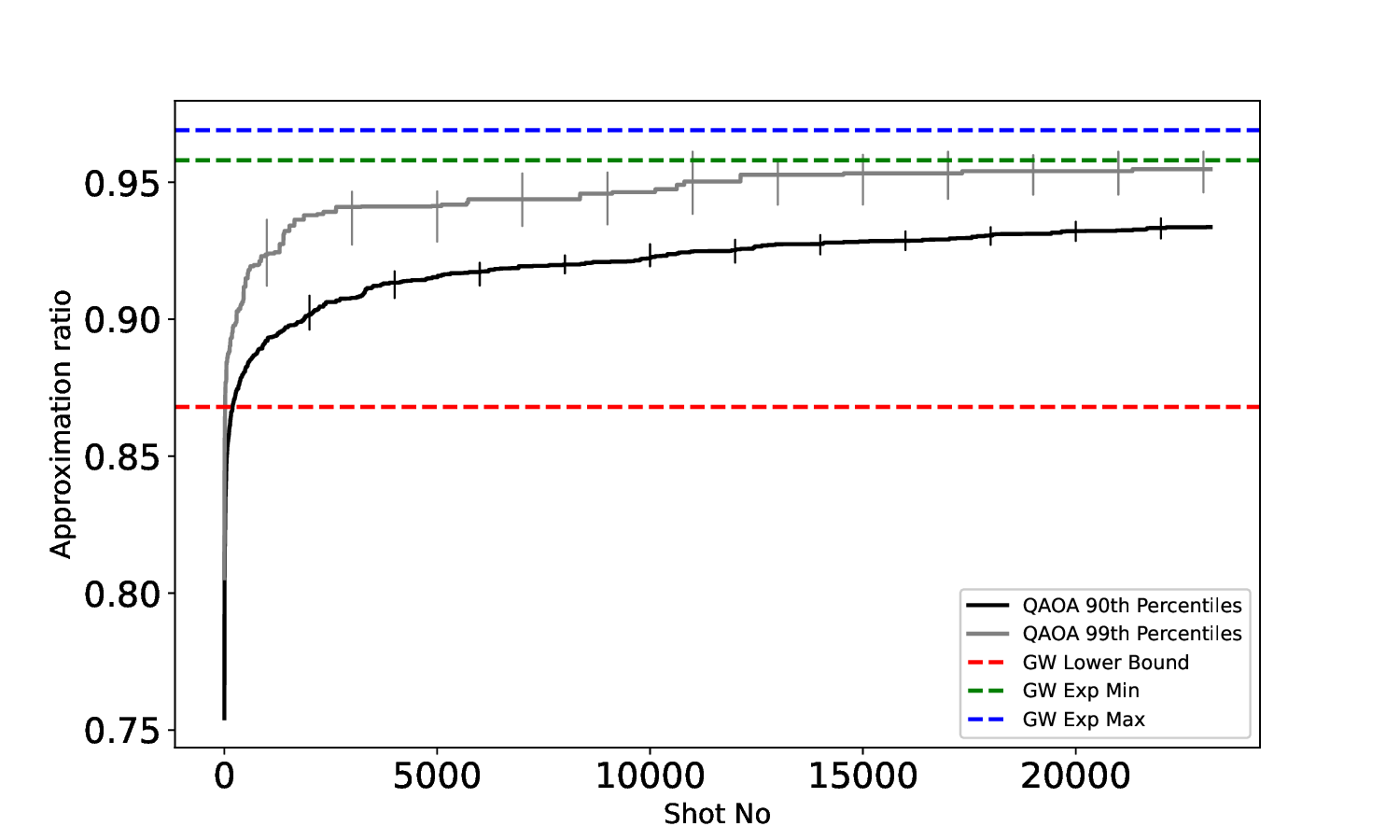}
    \caption{Aggregate performance across all instances. The figure shows the empirical $P_{90}(s)$ (black line) and $P_{99}(s)$ (gray line) percentiles as functions of the shot index $s$. These curves are compared with the GW lower bound (red dashed line), as well as the minimal (green dashed line) and maximal (blue dashed line) GW expectation values observed across the set of instances. The results indicate that, on average, fewer than $1\%$ of QAOA runs produce solutions whose quality reaches the level expected from the GW algorithm.}
    \label{fig:ba_n-29_m-9_132_gw_classical}
\end{figure}

\FloatBarrier

These findings should not be interpreted as inherent limitations of QAOA as an algorithmic paradigm. Rather, they highlight the performance gap that persists between current black-box implementations and a well-optimized classical baseline on small but non-trivial Max-Cut instances. Bridging this gap requires enhanced parameter-setting strategies, problem-aware ansätze and more efficient shot allocation methods.

\section{Conclusion}
In this work, we have introduced a shot-based benchmarking framework for evaluating specific implementations of QAOA for the Max-Cut problem. The proposed methodology adopts a black-box evaluation perspective that reflects the experience of typical end users by foregoing manual parameter tuning and instead relying on default initialization values or those recommended by the implementation itself.

We have also outlined an instance selection strategy tailored for benchmarking purposes. The generated graph instances, while limited in size, are designed to emulate structural features of real-world graphs. These instances are constructed to remain neutral with respect to algorithmic biases, maintaining sufficient complexity to distinguish between high- and low-quality solutions.

To demonstrate the applicability of our framework, we applied it to compare a specific QAOA implementation against the GW algorithm. Our analysis shows that, under this implementation, GW significantly outperforms QAOA. For users without parameter-tuning expertise, GW algorithm remains decisively superior at this scale. However, this finding is not necessarily an evidence against the potential of QAOA to achieve quantum advantage in solving Max-Cut. Rather, it highlights the limitations of the selected initialization and parameter update strategies in the evaluated implementation. The proposed benchmarking scheme offers a robust and generalizable tool for assessing QAOA performance and may aid in guiding the development of more effective algorithmic designs.

The methodology outlined in the previous subsections can be extended to other combinatorial optimization problems beyond QAOA for Max-Cut. To generalize this approach, one must first select a suitable combinatorial optimization problem. Alongside this, an appropriate variational quantum algorithm implementation should be identified, tailored to the chosen problem. A classical algorithm must also be chosen as a benchmark, ensuring it is a well-established or state-of-the-art approach commonly used for solving this type of problem.

A critical step in this process is the selection of problem instances. The dataset should consist of small instances that are computationally feasible for quantum algorithms but still complex enough to differentiate solution quality meaningfully. It is essential to avoid synthetic instances that are explicitly designed to favor or disfavor a particular algorithm, as this could lead to misleading conclusions. Likewise, instances with specific structures or regularities should be avoided since they may introduce biases that do not reflect real-world problem distributions. Ideally, the selected instances should be representative of practical optimization problems but scaled down to a size that remains tractable for either already existing or near-term quantum hardware.

Once the problem instances are selected, the quantum algorithm should be executed repeatedly to collect a sufficient amount of performance data. This allows for a statistical evaluation of its results, following the methodology outlined in previous subsections. The classical counterpart should be run under similar conditions, ensuring a fair comparison. Performance metrics such as approximation ratios, solution quality, and computational efficiency should be analyzed, taking into account theoretical expectations and performance guarantees for classical methods.

\bmhead{Funding} This research was funded by the Research Council of Norway through project number 332023. DOT also received support from the Estonian Research Council under grant PRG946 ``Secure Quantum Technology'' and the Center of Excellence ``Foundations of the Universe'', work group  ``Quantum Information and Computing'', TK202U7. 

\bmhead{Acknowledgements} The authors wish to thank Terje Nilsen at Kongsberg Discovery for the access to an H100 GPU for the simulations.



\begin{appendices}
\renewcommand{\thefigure}{\arabic{figure}}
\setcounter{figure}{4}
\renewcommand{\thetable}{\arabic{table}}

\section{Goemans-Williamson statistics}\label{appendix:gwstats}
In addition to comparing QAOA performance against the GW lower bound and expectation, it is also beneficial to track the actual empirical performance of the GW algorithm. Given that each random hyperplane sampling in GW is independent, let \( K \) denote the total number of samplings performed after solving the SDP relaxation. This allows for the collection of empirical statistics on the expected number of samples required to achieve a cut cost of at least \( C \).  

For a given graph and a target cost \( C \), define \( S_C \) as the number of samples that yield a cut with a cost of at least \( C \), i.e.,  
\begin{equation*}
S_C = \sum_{k=1}^{K} \mathbf{1}(C_k \geq C),   
\end{equation*}  
where \( \mathbf{1}(\cdot) \) is the indicator function and \( C_k \) denotes the cut cost obtained in the \( k \)-th sampling.  

The empirical expectation for the number of samples required to obtain a cut cost of at least \( C \) is then given by  
\begin{equation*}
\mathbb{E}_K[C] = \frac{K}{S_C}.   
\end{equation*}

Furthermore, given the exact Max-Cut value \( C_{\max} \), each cut cost \( C \) and \( C_k \) can be normalized by scaling it relative to \( C_{\max} \) and defining the approximation ratios as \( \alpha = C/C_{\max} \) and \( \alpha_k = C_k/C_{\max} \) respectively. Using this normalization, the empirical expectation for the number of samples required to achieve a cut with an approximation ratio of at least \( \alpha \) can be expressed as  
\begin{equation*}
\mathbb{E}_K[\alpha] = \frac{K}{S_\alpha}.    
\end{equation*}

\section{Instance generating scheme}

\subsection{Generating restrictions}\label{appendix:generatingrestrictions}
For the instance generation procedure, an upper bound was imposed on the GW expectation \( \mathbb{E}[\alpha_{\operatorname{GW}}]\) for each graph, ensuring that  
\begin{equation*}
\mathbb{E}[\alpha_{\operatorname{GW}}] \leq 0.97.   
\end{equation*}
This threshold accounts for the relatively small graph size, specifically \( |V| = 29 \), which inherently leads to high GW performance. Imposing this constraint ensures that the gap between the optimal solution and GW performance remains sufficiently large for meaningful benchmarking.  

To minimize the probability of obtaining high-quality solutions through random cut sampling, the following constraint was applied: the 99.9th percentile of all possible cuts remains below the GW lower bound \( \alpha_{\operatorname{GW}} \cdot C_{\max} \). Additionally, to ensure that the instances remain challenging for the GW algorithm and unfeasible for random sampling, the probability of observing a cut \( C \) that exceeds the GW expectation was set to be less than $2^{-|V|+1+7}=2^{-21}$. This corresponds to 
\begin{equation*}
29 \leq | \{ C \mid C > \mathbb{E}[C_{\operatorname{GW}}] \}| < 2^7.   
\end{equation*}

\subsection{Naming scheme}\label{appendix:instance-naming}
Each generated graph instance is assigned a filename based on its generation method and associated parameters. The naming scheme encodes key instance characteristics, facilitating easy identification and retrieval of datasets. Three distinct random graph models from the \texttt{networkx} Python library are utilized for instance generation: Connected Watts-Strogatz (CWS), Barabási-Albert (BA) and Erdős-Rényi (ER).
The filename structure follows a standardized pattern.

\begin{table}[h]  
    \centering
    \caption{Instance Naming Conventions for Different Graph Models} 
    \begin{tabular}{|c|c|c|c|}  
        \hline  
        \textbf{Graph Model} & \textbf{Parameters} & \textbf{Filename Format} \\  
        \hline  
        CWS & \( n, k, p \) & \texttt{cws\_n-\$n\_k-\$k\_p-\$p\_\$task\_id.gml} \\  
        \hline  
        BA & \( n, m \) & \texttt{ba\_n-\$n\_m-\$m\_\$task\_id.gml} \\  
        \hline  
        ER & \( n, p \) & \texttt{er\_n-\$n\_p-\$p\_\$task\_id.gml} \\  
        \hline  
    \end{tabular}       
\end{table}
\FloatBarrier

Here, \( n \) represents the number of nodes, while \( k \), \( p \), and \( m \) correspond to model-specific parameters. The integer variable \texttt{task\_id} is included to distinguish instances but does not encode any meaningful information about the graph parameters.

\section{Full list of experiments for the instance ba\_n-29\_m-9\_132}\label{appendix:experiments}

\begin{figure}[h]
    \includegraphics[width=\textwidth]{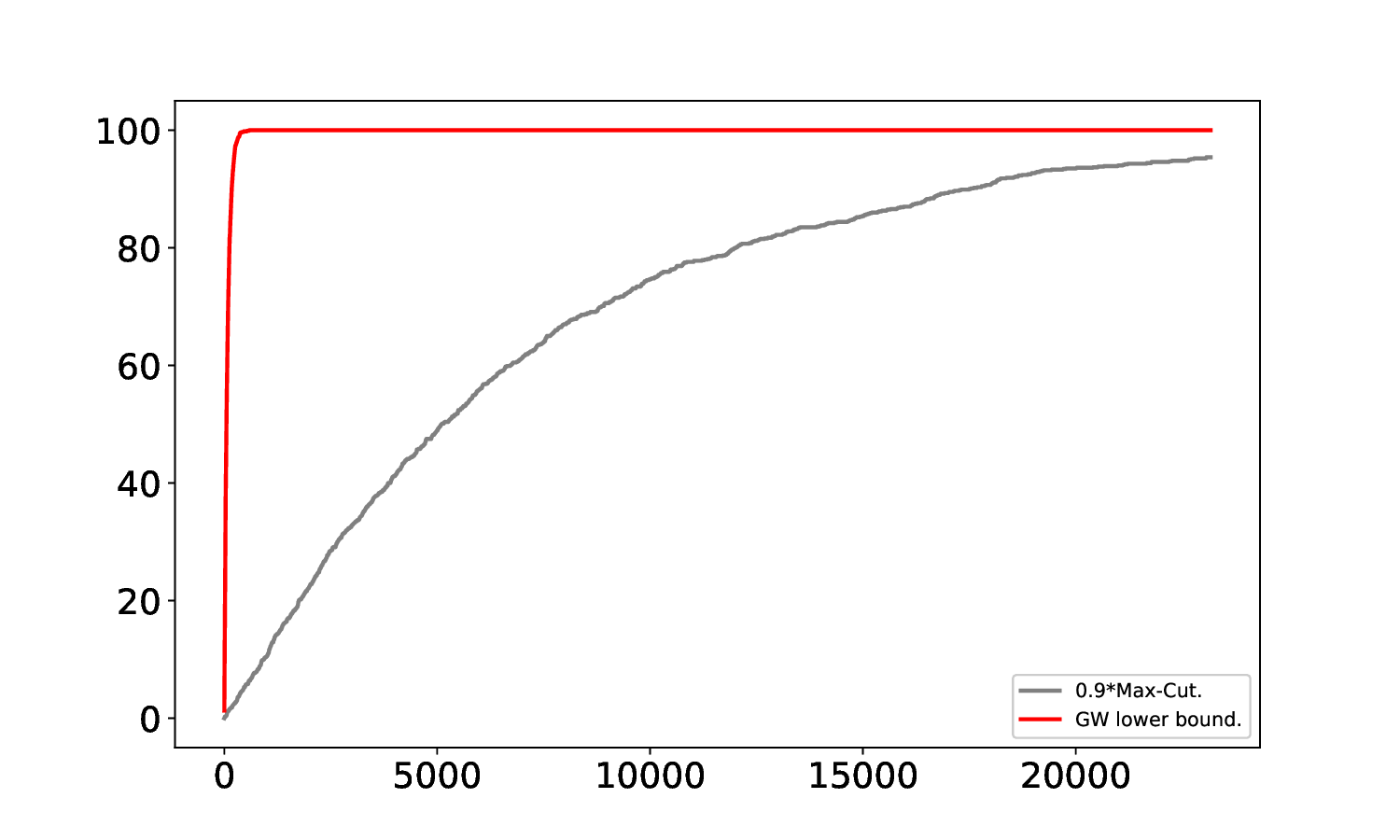}
    \caption{The percentage ($\%$) of the total number of QAOA runs (1000) as a function of the number of shots \(s \in \{1, 2, \ldots, \lfloor 2^{29/2} \rfloor \}\) for which at least a single QAOA output exceeds the GW lower bound (red line) and  \( 0.9 \cdot C_{\max} \) (gray line).}
    \label{fig:90maxcut}
\end{figure}

\begin{figure}[h]
    \includegraphics[width=\textwidth]{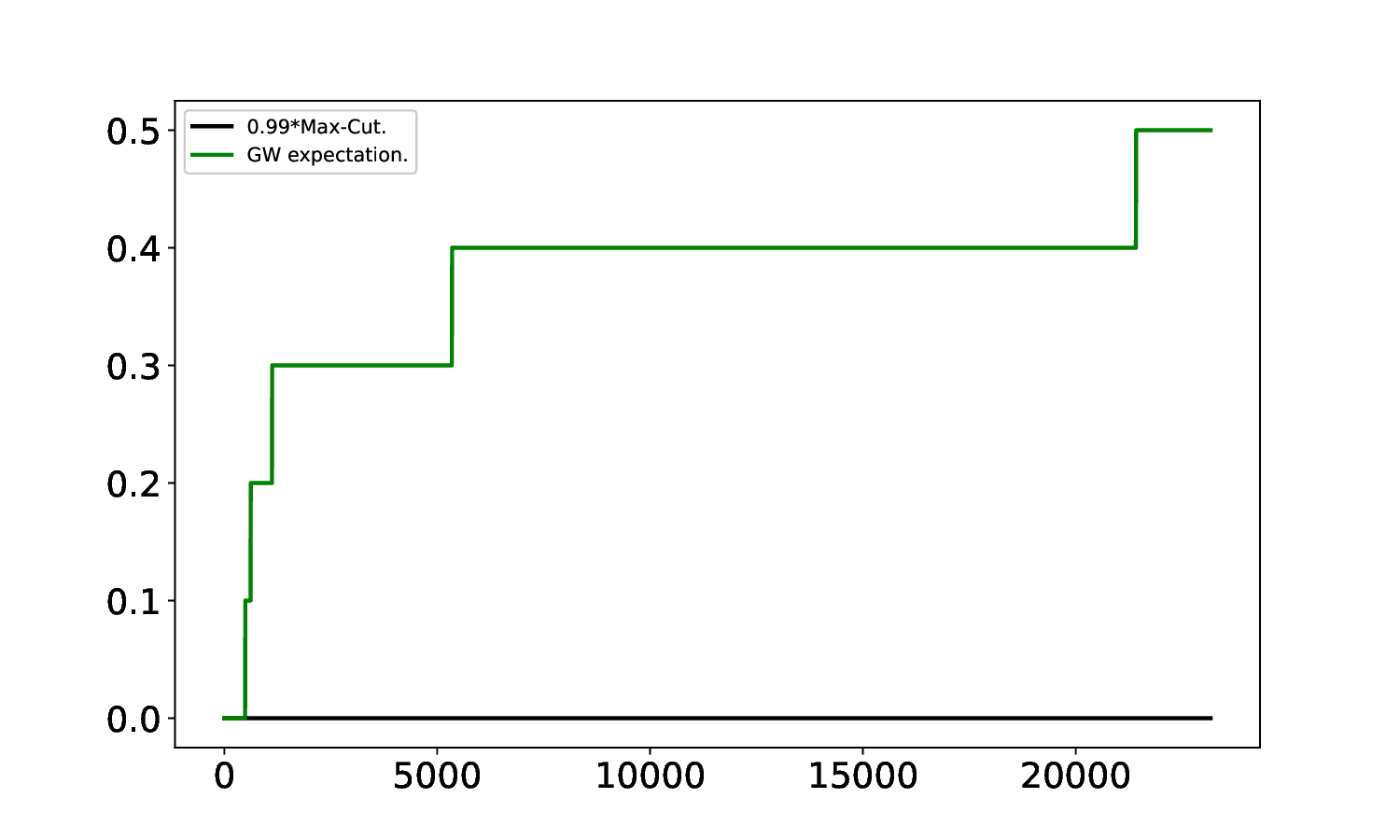}
    \caption{The percentage ($\%$) of the total number of QAOA runs (1000) as a function of the number of shots \(s \in \{1, 2, \ldots, \lfloor 2^{29/2} \rfloor \}\) for which at least a single QAOA output exceeds the GW expectation (green line) and \( 0.99 \cdot C_{\max} \) (black line).}
    \label{fig:99maxcut}
\end{figure}

\begin{figure}[h]
    \includegraphics[width=\textwidth]{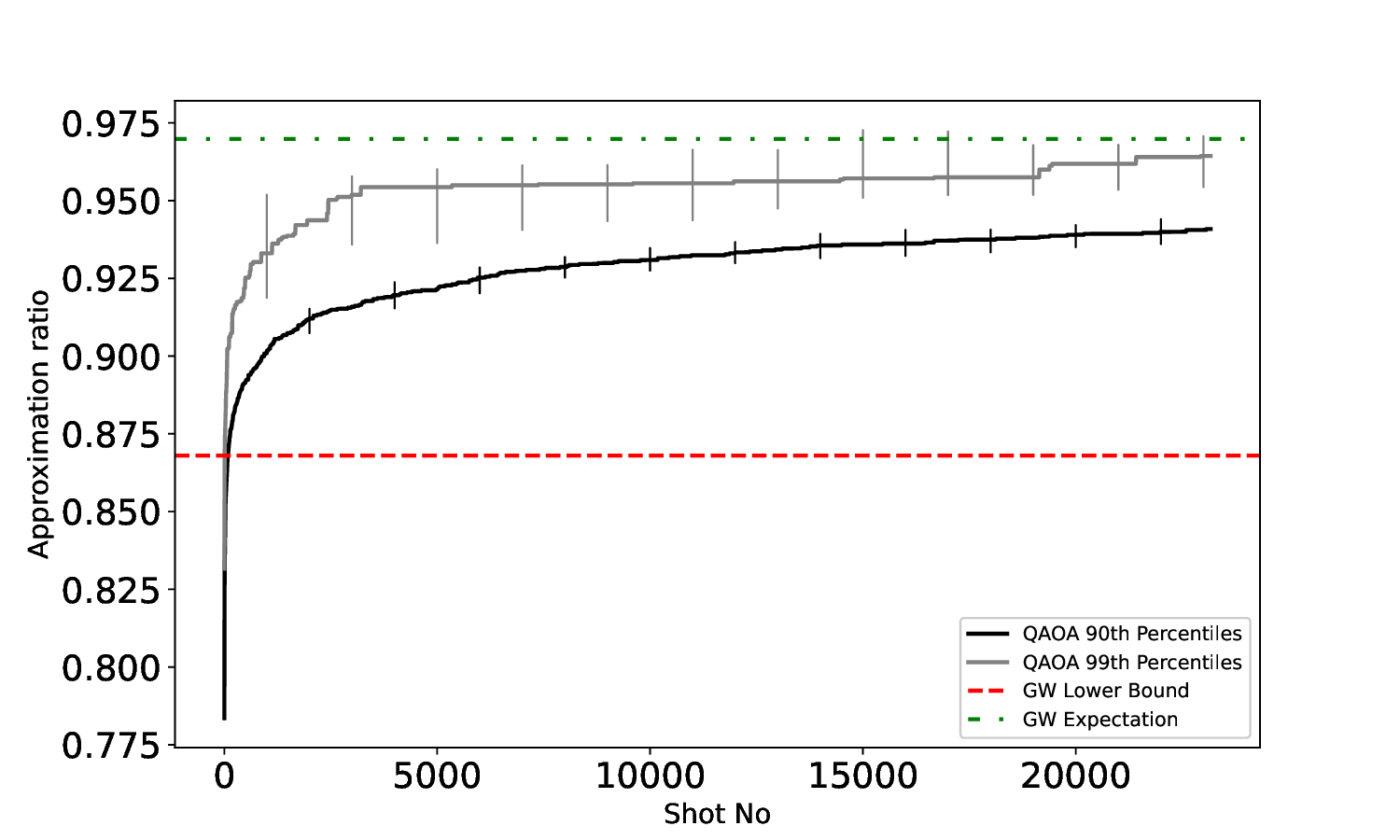}
    \caption{Comparison of \( P_{90}(s) \) (black line) and \( P_{99}(s) \) (gray line) values against the GW lower bound (red dashed line) and the GW expectation (green dash-dotted line) as a function of shot number \(s \in \{1, 2, \ldots, \lfloor 2^{29/2} \rfloor \}\).}
    \label{fig:ba_n-29_m-9_132_gw}
\end{figure}

\begin{figure}[h]
    \includegraphics[width=\textwidth]{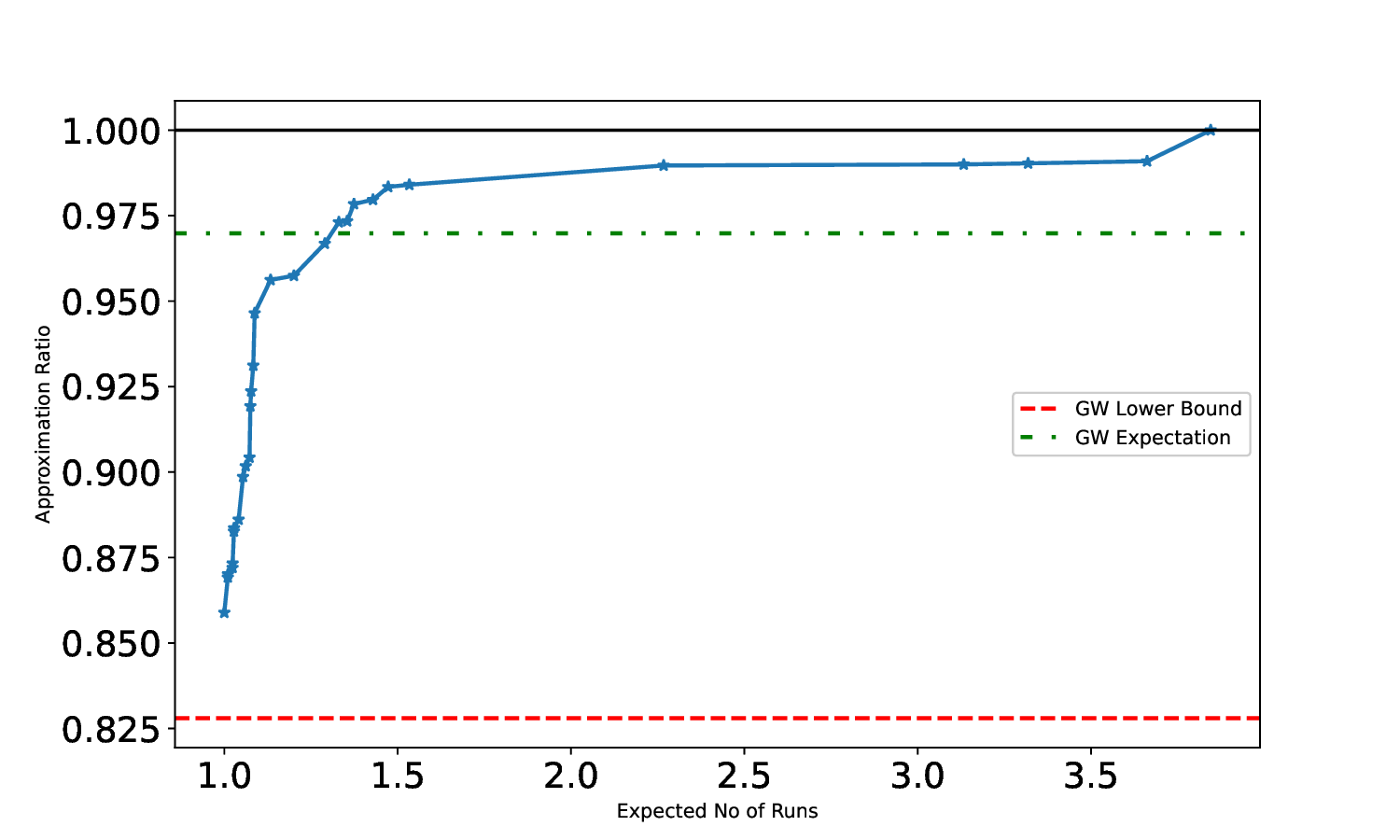}
    \caption{The plot depicts empirical expected numbers of GW runs ($\mathbb{E}_K[\alpha]$ for \(K=10^8\)) required for the GW algorithm to achieve a certain approximation ratio. GW lower bound (red dashed line) and \(\mathbb{E}[\alpha_{\operatorname{GW}}]\) (green dash-dotted line) are added for the reference.}
    \label{fig:ba_n-29_m-9_132_gw_classical}
\end{figure}

\FloatBarrier




\end{appendices}


\bibliography{sn-bibliography}

\end{document}